# High Mobility two-dimensional electron gas in modulation-doped β-(Al$_x$Ga$_{1-x}$)$_2$O$_3$/Ga$_2$O$_3$ heterostructures

Yuewei Zhang,[1,a)] Adam Neal,[2] Zhanbo Xia,[1] Chandan Joishi,[1,3] Yuanhua Zheng,[4] Sanyam Bajaj,[1] Mark Brenner,[1] Shin Mou,[2] Donald Dorsey,[2] Kelson Chabak,[5] Gregg Jessen,[5] Jinwoo Hwang,[6] Joseph Heremans,[4,6,7] and Siddharth Rajan[1,6,a)]

[1] Department of Electrical and Computer Engineering, The Ohio State University, Columbus, Ohio, 43210, USA

[2] Air Force Research Laboratory, Materials and Manufacturing Directorate, Wright-Patterson Air Force Base, Ohio 45433, USA

[3] Department of Electrical Engineering, Indian Institute of Technology Bombay, Mumbai 400076, India

[4] Department of Mechanical and Aerospace Engineering, The Ohio State University, Columbus, Ohio 43210, USA

[5] Air Force Research Laboratory, Sensors Directorate, Wright-Patterson Air Force Base, Ohio 45433, USA

[6] Department of Materials Science and Engineering, The Ohio State University, Columbus, Ohio, 43210, USA

[7] Department of Physics, The Ohio State University, Columbus, Ohio 43210, USA

**Abstract: Beta-phase Ga$_2$O$_3$ has emerged as a promising candidate for a wide range of device applications, including power electronic devices, radio-frequency devices and solar-blind photodetectors. The wide bandgap energy and the predicted high breakdown field, together with the availability of low-cost native substrates, make β-Ga$_2$O$_3$ a promising material compared to other conventional wide bandgap materials, such as GaN and SiC. Alloying of Al with β-Ga$_2$O$_3$ could enable even larger band gap materials, and provide more flexibility for electronic and optoelectronic device design. In this work, we demonstrate a high mobility two-dimensional electron gas (2DEG) formed at the β-(Al$_x$Ga$_{1-x}$)$_2$O$_3$/Ga$_2$O$_3$ interface through modulation doping. Shubnikov-de Haas oscillation was observed for the first time in the modulation-doped β-(Al$_x$Ga$_{1-x}$)$_2$O$_3$/Ga$_2$O$_3$ structure, indicating**

---

a)   Authors to whom correspondence should be addressed.
     Electronic mail: zhang.3789@osu.edu, rajan@ece.osu.edu

**a high-quality channel formed at the heterojunction interface. The formation of the 2DEG channel was further confirmed by a weak temperature-dependence of the carrier density, and the peak low temperature mobility was found to be 2790 cm$^2$/Vs, which is significantly higher than can be achieved in bulk-doped β-Ga$_2$O$_3$. The demonstrated modulation-doped β-(Al$_x$Ga$_{1-x}$)$_2$O$_3$/Ga$_2$O$_3$ structure lays the foundation for future exploration of quantum physical phenomena as well as new semiconductor device technologies based on the β-Ga$_2$O$_3$ material system.**

Beta-phase gallium oxide (β-Ga$_2$O$_3$) is a promising candidate for electronic device applications because of the large bandgap energy (4.7 eV) and the expected high breakdown field of 8 MV/cm.[1] Significantly, β-Ga$_2$O$_3$ is the first wide band gap material that can be grown from the melt, which makes it feasible to achieve large area bulk substrates on a manufacturable scale.[2-6] The good transport properties (mobility > 200 cm$^2$/Vs, and saturation velocity of ~ 2×10$^7$ cm/s)[1,7-9] make β-Ga$_2$O$_3$ very promising for a range of technological applications including high power electronics, detectors, and high frequency transistors. Further band gap tunability can be realized through the introduction of In and Al into β-Ga$_2$O$_3$, leading to β-(In,Ga)$_2$O$_3$ and β-(Al,Ga)$_2$O$_3$ alloys. β-(Al,Ga)$_2$O$_3$ is expected to be stable in the monoclinic phase over a broad range of compositions, and the band gap can be tuned from the band gap of β-Ga$_2$O$_3$ (4.7 eV) to above 6 eV[10,11]. This enables the realization of several semiconductor heterostructure designs such as modulation-doped electron channels, quantum wells and superlattices in this semiconductor system. Notably, unlike the other wide band gap semiconductor heterostructure system (III-Nitrides), heterostructures of monoclinic β-Ga$_2$O$_3$ do not exhibit spontaneous or piezoelectric polarization.

Recent efforts have led to the demonstration of various device structures with excellent performance, including Schottky diodes[12,13], metal-oxide-semiconductor field effect transistors (MOSFETs)[1,14-21] and metal-semiconductor field effect transistors (MESFETs)[1]. Experimental observations of high breakdown fields above 5 MV/cm have been reported for vertical Schottky diodes[12], and 3.8 MV/cm for lateral MOSFET transistors[16], which have already surpassed the material limit for GaN and SiC. While excellent device performance has been demonstrated using homoepitaxial β-Ga$_2$O$_3$ device structures, the β-(Al$_x$Ga$_{1-x}$)$_2$O$_3$/Ga$_2$O$_3$ heterojunctions have remained under-explored.

Preliminary demonstration of modulation-doped field effect transistors (MODFETs) using β-(Al$_x$Ga$_{1-x}$)$_2$O$_3$/Ga$_2$O$_3$ heterostructures have been reported.[22-24] Sheet charge densities above 5×10$^{12}$ cm$^{-2}$ were measured based on either Si-delta doping[22] or Ge doping[23] in the β-(Al$_x$Ga$_{1-x}$)$_2$O$_3$ layer, but the presence of parallel conduction through the low mobility channel in the (Al$_x$Ga$_{1-x}$)$_2$O$_3$ layer compromised the transport properties of the 2DEG. In this study, we show a direct evidence of a quantum confinement of electrons at the β-(Al$_x$Ga$_{1-x}$)$_2$O$_3$/Ga$_2$O$_3$ interface based on temperature-dependent Hall measurements and Shubnikov-de Haas (SdH) oscillations. We demonstrate room temperature mobility of 180 cm$^2$/Vs and low temperature mobility of 2790 cm$^2$/Vs achieved using the modulation-doped structure.

Figure 1(a) shows the β-(Al$_x$Ga$_{1-x}$)$_2$O$_3$/Ga$_2$O$_3$ MODFET structure used in this work. The samples were grown on a (010)-oriented Fe-doped semi-insulating β-Ga$_2$O$_3$ substrate using oxygen plasma-assisted molecular beam epitaxy (PA-MBE).[25,26] It consists of an unintentionally doped (UID) β-Ga$_2$O$_3$ buffer layer, 4.5 nm β-(Al$_x$Ga$_{1-x}$)$_2$O$_3$ spacer, a Si delta-doped layer, and 22.5 nm β-(Al$_x$Ga$_{1-x}$)$_2$O$_3$ cap layer. The growth details could be found in the supplementary information. Two samples were compared in this study, with the only difference being the UID β-Ga$_2$O$_3$ buffer layer thickness, which is 130 nm and 360 nm for sample A and B, respectively. The (Al$_x$Ga$_{1-x}$)$_2$O$_3$ layer thickness and the Al composition were estimated to be 27 nm and 18%, respectively, for both samples, based on high resolution XRD measurements of the (020) diffraction (Fig. 1(b)).[27] The observed diffraction fringes indicate sharp heterointerfaces. Both samples showed smooth surfaces with RMS roughness of ~ 0.45 nm obtained from AFM measurements as shown in the inset of Fig. 1(b).

The energy band diagram of the MODFET structure was obtained based on a self-consistent solution of the Schrodinger-Poisson equation assuming a conduction band offset (ΔE$_C$) of 0.4 eV, a surface depletion barrier of 1.4 eV, and a back-depletion due to Fe-doped semi-insulating substrate (assuming Fermi level pinned at the midgap), as shown in Fig. 1(c). A 2DEG is expected to form at the $(Al_xGa_{1-x})_2O_3/Ga_2O_3$ interface. When a donor concentration of $4.7\times10^{12}$ cm$^{-2}$ was adopted in the delta-doped layer, the simulated 2DEG density increased from $1.12\times10^{12}$ cm$^{-2}$ to $1.50\times10^{12}$ cm$^{-2}$ as the buffer layer was increased from 130 nm to 360 nm, because of a corresponding reduction of backside depletion.

To achieve ohmic contact to the channel, a contact regrowth technique was developed using $SiO_2$ as a regrowth mask. The contact regrowth and device fabrication processes are described in supplementary information. Ohmic contacts were verified by transfer length measurements, with extracted contact resistances of 9.3 $\Omega$ mm for sample A and 4.1 $\Omega$ mm for sample B, respectively, which were limited by the sidewall contacts between the regrown contacts and the low charge density 2DEG channel.

Temperature-dependent Hall measurements were carried out using a Van der Pauw structure, as shown in Fig. 2. Both samples showed weak temperature dependence in the measured carrier density, which dropped from $1.12\times10^{12}$ cm$^{-2}$ to $1.07\times10^{12}$ cm$^{-2}$ in sample A and from $2.25\times10^{12}$ cm$^{-2}$ to $2.05\times10^{12}$ cm$^{-2}$ in sample B upon lowering the temperature. This is in contrast with carrier freeze-out in bulk-doped β-$Ga_2O_3$ at low temperatures[28], and serves as a direct proof of a degenerate electron gas with no parallel conduction in the $(Al_xGa_{1-x})_2O_3$ barrier layer. Using a higher Si sheet density in the delta-doped layer led to partial freeze-out of charge at low temperatures, which could be attributed to a parallel conduction channel in the barrier layer. At present, the maximum charge density that can be confined in the adopted $(Al_{0.18}Ga_{0.82})_2O_3/Ga_2O_3$ MODFET structure without introducing a parasitic channel is estimated to be approximately $2\times10^{12}$ cm$^{-2}$. Further increase in the 2DEG charge density requires higher conduction band offset or using a relatively thinner spacer layer.

The room temperature mobility was measured to be 162 cm$^2$/Vs and 180 cm$^2$/Vs for sample A and B, respectively. Both samples showed a sharp increase in the Hall mobility with decreasing temperature, with a peak mobility of 990 cm$^2$/Vs at 60 K for sample A and 2790 cm$^2$/Vs at 50 K for sample B. These are significantly higher than the highest reported mobility values obtained in bulk-doped β-$Ga_2O_3$, an expected benefit of the spatial separation between the impurities and the modulation-doped 2DEG channel. Their mobility values dropped off slightly upon further lowering of the measurement temperature. Similar phenomenon has been observed in early works on modulation-doped AlGaAs/GaAs transistors[29], and was attributed to impurity scattering. To understand the scattering mechanisms that are causing the differences between the samples, the temperature dependence of electron mobility was analyzed by considering various scattering mechanisms (discussed in supplementary materials).

The measured and calculated mobility results are shown in Fig. 2. At low temperature, the remote impurity scattering limited mobility is estimated to be significantly above the measured values, indicating that remote impurity scattering is not a limiting mechanism for the studied MODFET structure. To fit the measured data, the interface roughness and the background impurity density were adjusted in the calculations. The vertical/ lateral displacements of the interface were assumed to be 0.45 nm/ 4.7 nm for both samples for the best fittings of the measured data. The extracted effective charged impurity density was estimated to be $1.2\times10^{18}$ cm$^{-3}$ in sample A, while it is $1.5\times10^{17}$ cm$^{-3}$ in sample B. The reduction of the background impurity density in sample B contributed to the notable mobility increase at low temperatures, and it is attributed to less impurity (such as Fe) diffusion from the substrate surface due to a thicker buffer layer growth. The residual background charge could also have contributions from native defects formed during the MBE growth[8], such as Ga vacancies ($V_{Ga}$), which are expected to be deep acceptors.[30] We note here that these estimates are based on monovalent charged impurities. Since Ga vacancies can be charged up to the 3+ state, the true background defect density (as opposed to *effective* defect density) could be lower than $1.5\times10^{17}$ cm$^{-3}$.

Even though an apparent increase in the low temperature mobility was achieved by increasing the buffer layer thickness, the electron scattering is dominated by polar optical phonon scattering in the high temperature range[8,9,31], leading to similar mobility values at room temperature for both samples. While longitudinal optical-plasmon coupling could lead to better screening of the optical phonon scattering and therefore higher electron mobilities at room temperature[8,9], this effect is not significant in the low carrier density range below $2 \times 10^{12}$ cm$^{-2}$, and was therefore not considered in the mobility calculations. Increasing the 2DEG density is necessary to take advance of the screening effects.

The high channel mobility at low temperature allowed for the measurement of Shubnikov-de Haas oscillations of the transverse magnetoresistance ($R_{xx}$) with varied magnetic field perpendicular to the sample surface. Both samples showed negative magnetoresistance at low magnetic field, attributed to weak localization[32-34], with SdH oscillations developing above four Tesla. The oscillation components of $R_{xx}$, with the background subtracted, are shown in Fig. 3(b) and (d) as a function of reciprocal magnetic field (1/B). Only one period of the oscillation was observed for sample A in the magnetic field range below 14 T, while multiple oscillations developed below 7 T for sample B at varied measurement temperatures benefitting from the higher channel mobility. The 2DEG concentration can be estimated based on the period of $\Delta(1/B)$ through the expression: $\Delta(1/B) = e/\pi\hbar n_{2D}$.[32] The determined 2DEG densities from SdH oscillation are $1.15 \times 10^{12}$ cm$^{-2}$ and $1.96 \times 10^{12}$ cm$^{-2}$ for sample A and B, respectively, which are consistent with low-field Hall measurements.

The obtained SdH oscillations in sample B make it possible to analyze both the effective mass of the 2DEG and the quantum scattering time ($\tau_q$) in β-Ga$_2$O$_3$. The temperature dependence of the SdH oscillation amplitude ($A$) at a fixed magnetic field ($B$) can be used to determine the effetive mass of the 2DEG. The oscillation amplitude is related to the measurement temperature ($T$) by[35]

$$\ln\left(\frac{A}{T}\right) = C_1 - \ln\left[\sinh\left(\frac{2\pi^2 m^*}{e\hbar B} k_B T\right)\right],$$

where $k_B$ is the Boltzmann constant, $C_1$ is a temperature-independent constant at a fixed magnetic field. The dependence of $\ln(A/T)$ on temperature and the fittings for the effective mass are plotted in Fig. 4(a) at three magnetic fields. The electron effective mass is extracted to be $m^* = 0.313 \pm 0.015$ $m_0$. While the electron transport is confined in the (010) plane, near isotropic effective mass was predicted[36] for β-Ga$_2$O$_3$ and the extracted effective mass is in close agreement with the theoretical calculations[36] and band structure measurements[37,38].

Following the estimation of the effective mass, the quantum scattering time ($\tau_q$) can be evaluated using a Dingle plot[39,40]. At a fixed temperature, the oscillation amplitude is related to the inverse of the magnetic field by[39,40]

$$\ln\left(\frac{1}{4}\frac{A}{R_0}\frac{\sinh(\chi)}{\chi}\right) = C_2 - \frac{\pi m^*}{e\tau_q}\frac{1}{B},$$

where, $\chi = 2\pi^2 m^* k_B T/\hbar e B$, $R_0$ is the zero field resistance, and $C_2$ is a constant that is independent of the magnetic field at a given temperature. Fig. 4(b) shows the dependence of $ln[(A/4R_0)(sinh(\chi)/\chi)]$ on $1/B$ extracted from the SdH oscillation at 3.5 K. Linear fitting of the experimental data gives a quantum scattering time of 0.33 ps. In comparison, the transport lifetime ($\tau_t$) was estimated to be 0.44 ps from the low field Hall mobility ($\tau_t = m^*\mu/e$) using the extracted effective mass of $0.313m_0$. The ratio between the transport lifetime and quantum scattering time is therefore $\tau_t/\tau_q \sim 1.5$. This is close to unity, and indicates that the electron scattering events are dominated by large angle scatterings, such as interface roughness scattering or background impurity scattering,[41] in agreement with the mobility calculations shown in Fig. 2(c).

To demonstrate the feasibility of device applications, modulation-doped field effect transistors were fabricated using a Pt/Au metal stack to define the Schottky gate contact. The output and transfer characteristics of three-terminal transistors are shown in Fig. 5. A maximum drain current of $I_{DS}$ = 46 mA/mm was obtained at $V_{DS}$ of 10 V, and $V_{GS}$ of 2 V. The transconductance ($g_m$) showed a peak of 39 mS/mm and dropped off at higher gate bias, which we attribute to the decrease of modulation efficiency due to charge transfer into the barrier layer.[42] The devices showed low output conductance after saturation, indicating good gate control of the 2DEG. $I_{DS}$ showed above 9 orders of magnitude rectification at $V_{DS}$ = 10 V, and the subthreshold slope is estimated to be 91 mV/decade. The extracted threshold voltage is 0.5 V, corresponding to normally-off operation under the Pt-gate. High frequency small-signal measurements on this device showed a cutoff frequency of 3.1 GHz, and maximum oscillation frequency of 13.1 GHz at $V_{DS}$ of 10 V and $V_{GS}$ of 1.5 V.

In summary, the formation of a high mobility 2DEG channel was achieved using modulation doping in a (010)-oriented β-$(Al_xGa_{1-x})_2O_3$/$Ga_2O_3$ heterostructure with Si delta-doping in the barrier layer. Temperature dependent Hall measurement showed nearly constant charge density in the temperature range of 5 K to 300 K. Both the room temperature mobility of 180 cm$^2$/Vs and low temperature peak mobility of 2790 cm$^2$/Vs exceeded the highest experimental mobility values for bulk β-$Ga_2O_3$. This is attributed to the spatial separation between ionized impurities and the 2DEG. The mobility calculations indicate that the low temperature mobility is limited by ionized impurity scattering, while the room temperature mobility is mainly limited by phonon scattering. The high mobility values allowed for the observations of the SdH oscillations at low temperatures. The 2DEG densities extracted from SdH oscillations are consistent with the obtained charge densities from low field Hall measurements. The effective mass in the (010) plane is estimated to be $m^* = 0.313 \pm 0.015\ m_0$ based on the SdH oscillations. The demonstration of a high-quality heterojunction and quantum transport in the β-$(Al_xGa_{1-x})_2O_3$/$Ga_2O_3$ wide band gap semiconductor system reported here lays the foundation for future investigation of the materials science, physics, and device applications of the monoclinic β-$(Al_xGa_{1-x})_2O_3$ semiconductor system.

**Acknowledgement:**

We acknowledge funding from the Office of Naval Research under Grant No. N00014-12-1-0976 (EXEDE MURI). The project or effort depicted was or is sponsored by the Department of the Defense, Defense Threat Reduction Agency (Grant HDTRA11710034). The material is partially based upon the work supported by the Air Force Office of Scientific Research under award number FA9550-18RYCOR098. The content of the information does not necessarily reflect the position or the policy of the federal government, and no official endorsement should be inferred. We acknowledge funding from The Ohio State University Institute of Materials Research (IMR) Multidisciplinary Team Building Grant.

**Figures:**

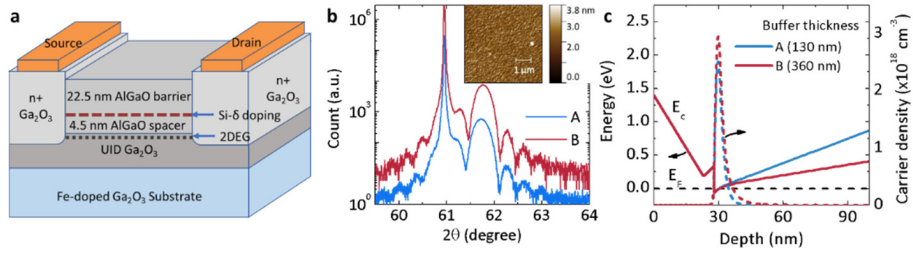

**Fig. 1 Epitaxial stack, XRD/AFM results and energy band diagram of the modulation-doped structures. a,** Schematic epitaxial stack of the MODFET structure. The UID buffer layer thicknesses are 130 nm and 360 nm in sample A and B, respectively. Here, AlGaO represents for $(Al_xGa_{1-x})_2O_3$. **b,** XRD of the (020) diffraction patterns and AFM image (inset) after growth. Both samples showed ~ 18% Al content. **c,** Equilibrium energy band diagram and calculated 2DEG charge distribution.

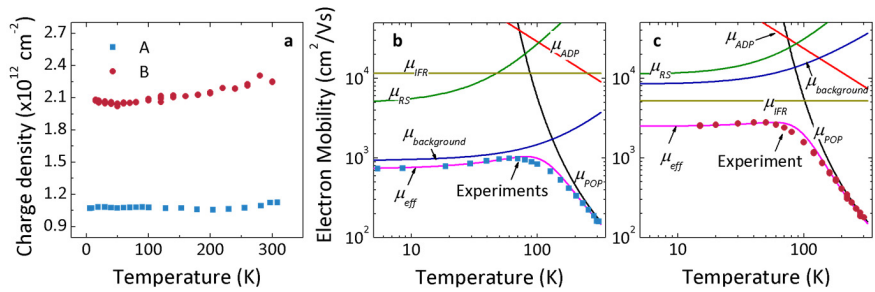

**Fig. 2 Charge density and electron mobility of the modulation-doped β-$(Al_xGa_{1-x})_2O_3$/$Ga_2O_3$ heterostructures. a,** Temperature-dependence of charge density measured using a Van der Pauw configuration. Both samples showed weak temperature dependence in the measured temperature range. **b-c,** Experimental and calculated electron mobilities for sample A (**b**) and sample B (**c**) by considering various scattering mechanisms, including: polar optical phonon scattering ($\mu_{POP}$), remote impurity scattering ($\mu_{RS}$), background impurity scattering ($\mu_{Background}$), interface roughness scattering ($\mu_{IFR}$), and acoustic deformation potential scattering ($\mu_{ADP}$). The β-$Ga_2O_3$ material parameters used in the calculations are: $m^*$=0.313$m_0$, static dielectric constant $\varepsilon_s$=10.2, high-frequency dielectric constant $\varepsilon_\infty$=3.57, sound velocity $v_s$=6800 m/s, mass density $\rho$=5880 kg/m³, acoustic deformation potential $\xi_{ADP}$=6.9 eV, polar optical phonon energy $\xi_{POP}$=44 meV.

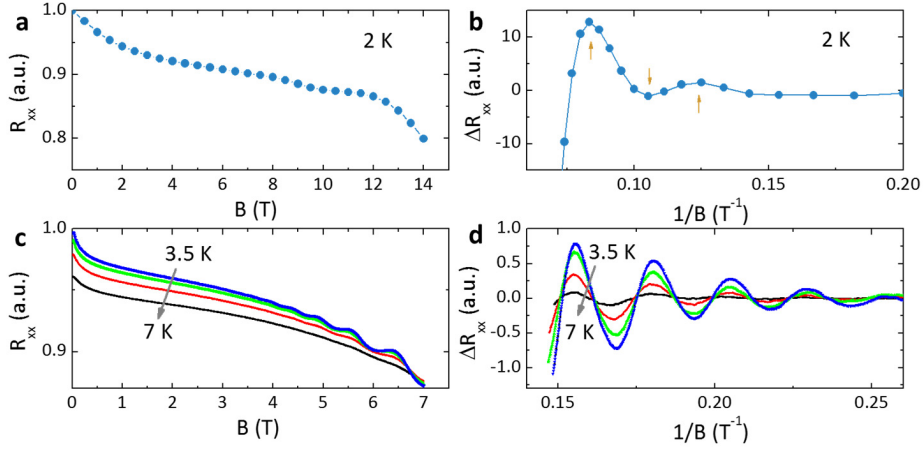

**Fig. 3 SdH oscillations of the transverse magnetoresistance measured with a magnetic field perpendicular to the sample surface.** Van der Pauw configuration was used for the SdH oscillation measurement. **a-b**, Dependence of $R_{xx}$ on B, and $\Delta R_{xx}$ on 1/B for sample A measured at 2 K. **c-d**, Dependence of $R_{xx}$ on B, and $\Delta R_{xx}$ on 1/B for sample B measured at 3.5, 4, 5, 7 K.

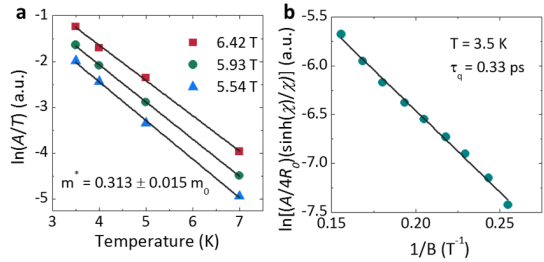

**Fig. 4 Fitting for the electron effective mass and quantum scattering time. a**, Fitting for the effective mass for three magnetic field values. The effective mass is estimated to be $m^* = 0.313 \pm 0.015\ m_0$. **b**, Dingle plot for the extraction of the quantum scattering time.

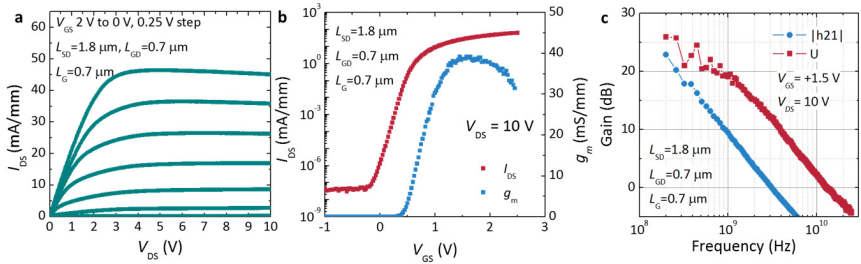

**Fig. 5 Output, transfer and RF characteristics of a MODFET transistor. a**, Output characteristics measured with gate bias $V_{GS}$ from 2 V to 0 V at a step of 0.25 V. **b**, Transfer characteristics measured under a drain bias of $V_{DS}$=10 V. **c**, RF characteristics measured at $V_{DS}$=10 V and $V_{GS}$=1.5 V. The gate length,

gate-drain spacing, source-drain spacing of the device are $L_G$=0.7 μm, $L_{GD}$=0.7 μm and $L_{SD}$=1.8 μm, respectively.